\documentclass[twocolumn,aps,prb,graphicx,epsfx,showpacs]{revtex4}%
\usepackage{amsmath}
\usepackage{amsfonts}
\usepackage{amssymb}
\usepackage{graphicx}%
\begin{document}
\title{On the optical conductivity  of Electron-Doped Cuprates I: Mott Physics}
\author{A. J.Millis$^{1}$, A. Zimmers$^{2,3}$, R. P. S. M.  Lobo$^{2}$, N. Bontemps$^{2}$}
\affiliation{$^{(1)}$Department of Physics, Columbia University\\
534 W. 120th St. New York , N. Y.  10027\\
$^{(2)}$LPS-ESPCI, CNRS UPR 5, 10 rue Vauquelin, 75231, Paris Cedex 5, France\\
$^{(3)}$ Center for Superconductivity Research\\ University of Maryland, College Park, MD  20742 }

\begin{abstract}
The doping and temperature dependent conductivity of electron-doped cuprates
is analysed. The variation of kinetic energy with doping is shown to imply
that the materials are approximately as strongly correlated as the hole-doped
materials. The optical spectrum is fit to a quasiparticle scattering model;
while the model fits the optical data well, gross inconsistencies with
photoemission data are found, implying the presence of a large,
strongly doping dependent Landau parameter.

\end{abstract}
\maketitle

\section{Introduction}

Despite 15 years of intensive study many properties of the cuprate
superconductors remain imperfectly understood. A crucial set of questions
involves  charge transport. Angle resolved photoemission
measurements \cite{Damascelli03} suggest the presence of reasonably well
defined electron-like quasiparticle excitations characterized by a Fermi
surface with position very close to that of band theory and a quasiparticle velocity only
moderately renormalized from the band value. 
Optical measurements  reveal a frequency dependent conductivity
\cite{Orenstein90,Uchida91} with a strong doping dependence
and an integrated low frequency absorption strength
("spectral weight") markedly smaller
than that predicted by band theory \cite{Millis04d}.

There is no generally accepted interpretation
of the measured conductivity of high-$T_c$ cuprates.  Some authors argue that it may be
understood in terms of quasiparticles scattered by a frequency and temperature
dependent scattering rate \cite{Schlesinger89,Orenstein90,Abrahams01}; others
that it should be understood in more exotic terms \cite{Anderson00,Marel03}.
Recently, the issue of the adequacy of a quasiparticle-only description has been reexamined
\cite{Millis03a}. For hole-doped materials at optimal doping a model involving
only quasiparticles with velocity and mean free path taken from angle-resolved photoemission
experiments, was  shown to be inconsistent with the data.  

A suppression of low frequency optical oscillator strength 
is  expected in materials, such as
the high $T_c$ superconductors, which  may be regarded as doped Mott insulators.
Our understanding of the physics of doped Mott insulators is far from complete.
However, the development over the last 15 years of the 'dynamical mean field
method'  has provided an important theoretical step forward \cite{Georges96},
providing a practical (and in many cases apparently reliable) method
for calculating properties of strongly correlated materials and leading to new
insights into the one-electron (photoemission and specific heat) properties
of doped Mott insulators \cite{Imada00}. A crucial assumption in this method
is that the electron self energy depends much more strongly on energy
than on momentum. This assumption may be theoretically justified
in a limit of infinite spatial dimensionality. It implies that 'vertex corrections'
may be neglected, making the calculation of the optical conductivity
straightforward once the electron self energy has been determined.  In other words,
the key consequence of this theoretical approach is that the 
Mott correlations are expressed mainly through the electron self energy,
and in particular the suppression of low frequency spectral weight occurring
as the Mott phase is approached is caused by a divergence of the electron effective mass.
An alternative class of theoretical approaches \cite{Ioffe89} involves ascribing the Mott
spectral weight suppression to a 'vertex correction 'which diverges as the insulating
phase is approached. 

The experimental status of these predictions is unsettled. 
Good agreement between dynamical mean field  calculations and data has
been found for electronically three dimensional compounds such as $V_2O_3$
\cite{Thomas95}.  However, the applicability of the method to two dimensional
correlated materials is not clear.  
The discrepancies between the quasiparticle-only model
and BSCCO optical data were interpreted by  \cite{Millis03a} 
as implying the presence of a relatively large
vertex correction, but the evolution of the vertex correction with doping was not determined
because only for optimally doped BSCCO were a consistent set of photoemission
and optical data available. One implementation of the vertex correction idea, 
the slave-boson-gauge theory approximation
\cite{Ioffe89} has  a reasonable qualitative correspondence
with the measured conductivity, but has been shown to make predictions for the 
doping and temperature dependence of the superconducting penetration depth
which disagree sharply with data \cite{Ioffe02a}.  

Recent experimental developments may offer a new perspective
on the charge dynamics of high temperature superconductors
and therefore of low dimensional Mott insulators.
Improvements in sample preparation have led to a systematic set of
measurements on \textit{electron-doped} cuprate materials
\cite{Onose01,Onose04,Zimmers04}. 
The electron-doped compounds, unlike the more extensively studied hole-doped compounds,
display at $x \le x=0.15$ and low enough temperature 
clear signatures of a density wave gap in the observed
conductivity \cite{Onose01,Onose04,Zimmers04}. Recent theoretical works 
\cite{Kusko02,Tremblay04} have explained the difference
in magnetic behavior between electron-doped and hole-doped compounds
in terms of a model in which the electron-doped compounds
exhibit   'Mott' correlations which are weaker than in the hole-doped ones,
and are doping dependent.

In this paper we undertake 
a systematic analysis of the observed
conductivity of the electron doped materials. 
Measurements at lower temperatures and lower dopings
reveal gap-like features which may be associated with
antiferromagnetic order \cite{Onose04,Zimmers04}.
We consider data
only at dopings and temperatures such that the density wave gap effects
do not affect the analysis; a companion paper reports results of more detailed
studies of density wave gap effects.
We show that the magnitude and
doping dependence of the kinetic energy implies that the
electron-doped cuprates are approximately as
stronly correlated as the hole-doped compounds. We further show that the
canonical model of quasiparticles scattered by a (possbily large) self energy
is an entirely inadequate description of the data.  
The rest of this paper is organized as follows. In section II we define the
model and the quantities of interest. Section III presents the specific 
form of the self energy used in our
detailed analysis. Section IV presents an analysis of the qualitative features
of the conductivity, in particular the kinetic energy and optical mass
enhancement. Section V presents an attempt to model the conductivity under the
'no-vertex-corrections' approximation using the self energy presented in
section III. Section VI outlines implications for the photoemission spectra
and section VII is a conclusion.

\section{Model}

\subsection{Overview}

The conventional description of the motion of electrons in solids is in terms
of electrons, moving with a dispersion defined by a band structure calculation
and modified by a self energy function expressing the effects of interactions
not included in the band calculation.
In this section we present the band structure which seems likely to
be relevant to the materials, and define the optical conductivity and related
quantities including the 'kinetic energy' which is a fundamental measure of
the correlation strength of the materials.

\subsection{Band Structure}

The canonical band structure of high-$T_{c}$ materials \cite{Andersen95} is
\begin{align}
\varepsilon_{p}  &  =-2t_{1}\left(  \cos p_{x}+\cos p_{y}\right)  +4t_{2}\cos
p_{x}\cos p_{y}\label{ep}\\
&  -2t_{3}\left(  \cos2p_{x}+\cos2p_{y}\right) \nonumber
\end{align}
The parameters $t_1,t_2,t_3$ have been obtained from band theory calculations
for {\it hole-doped} compounds
by a 'down-folding' procedure \cite{Andersen95}; canonical values are
\begin{align}
t_{1}  &  =0.38eV\label{t1}\\
t_{2}  &  =0.32t_{1}\label{t2}\\
t_{3}  &  =0.5t_{2} \label{t3}%
\end{align}
The Fermi surface found in band theory calculations
performed directly on electron-doped materials
 \cite{Massida89} is almost identical to the Fermi surface
following from Eq \ref{ep}. 

\begin{figure}[ptb]
\includegraphics[width=3.0in]{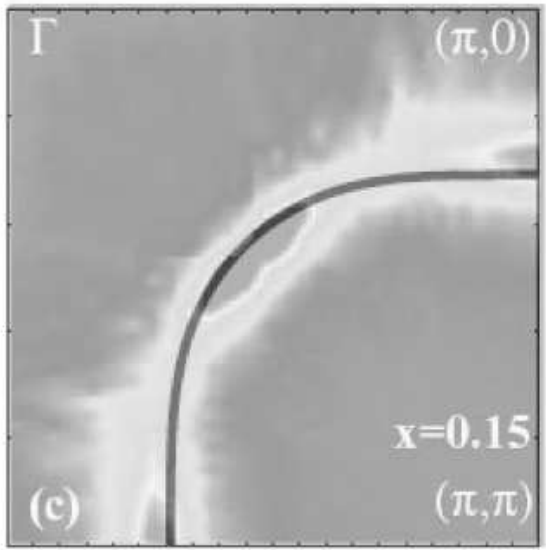}
\caption{\textit{Solid line: Fermi surface of electron doped cuprates at $x=0.17$ calculated 
from Eq  \ref{ep} using the 
standard band parameters given in the text and overlaid on a 
false-color representation of the near-chemical-potential photoemission
intensity \cite{Armitage02}  derived from angle resolved photoemission
measurements  of an NCCO sample with  nominal electron-doping $x=0.15$.
In this measurement the Fermi surface is located in the light-shaded region.
 }}%
\label{fig:Fermi surface}
\end{figure}

For the energy dispersion implied by these parameters, at all relevant carrier
concentrations the Fermi surface is hole-like (roughly circular, centered at
the $\pi,\pi$ point).  The Fermi surface implied by
Eq \ref{ep} for  electron-doping of $x=0.17$ is shown as the solid line in Fig
\ref{fig:Fermi surface} overlaid on a false-color
representation of the experimental near-fermi-surface
photoemission intensity \cite{Armitage02}. In this measurement
the fermi surface  is in the lighter-shaded.  (Other 
very recent measurements have found a slightly different shape \cite{Claesson04},
the differences are too small to be relevant to the present paper). 

Note that although the correspondence between the
calculated and measured Fermi surfaces is not perfect, it is reasonably
good. Note also that  the area enclosed
by the calculated curve is equal to that enclosed by the measured one, suggesting
that the actual doping of the region measured in the experiment is slightly higher
than the nominal doping of $x=0.15$.
The strong similarity between the calculation and the data appears
to rule out the possibility  of gross differences in the underlying Fermiology
between hole-doped to electron-doped materials, in particular contradicting
the theoretical suggestion \cite{renormbandparam} that many-body physics effects could lead to 
a change in sign of $t_2$ between electron and hole-doped materials.
We believe that the agreement between the observed Fermi surface and the
band theory one justifies the use of the tight-binding parameters in modelling the optical data. 

\begin{figure}[ptb]
\includegraphics[width=3.0in]{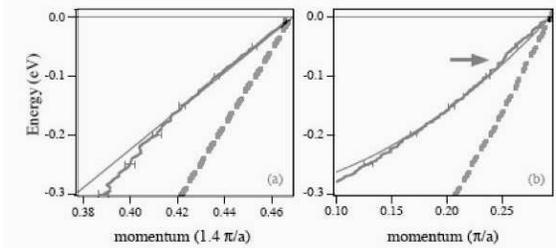}
\caption{\textit{Observed quasiparticle dispersions (solid lines) 
from angle resolved photemission measurements on
$Nd_{1.85}Ce_{0.15}CuO_4$  \cite{Armitage02} compared to tight binding 
model prediction (dashed lines).
Panel a: zone diagonal $((0,0) \rightarrow (\pi,\pi))$; Panel (b) 
zone face $((\pi,0) \rightarrow (\pi,\pi))$.
 }}%
\label{fig:dispersion}
\end{figure}

In wide classes of 'correlated electron' materials, standard band theory calculations
produce Fermi surfaces in reasonable agreement with experiment; however, electronic
dispersions are often substantially renormalized. Fig \ref{fig:dispersion} compares the
dispersion obtained from angle-resolved photoemission measurements along
two high-symmetry directions in the Brillouin zone, to those obtained
from the simple tight binding parameters used above. One sees that
in both symmetry  directions the measured velocities are roughly
half of the band velocities.

\subsection{Optical conductivity}

The optical conductivity $\sigma(\Omega)$ is the response function relating
current to applied uniform, transverse electric field; it is given by

\begin{equation}
\sigma(\Omega)=\frac{e^{2}}{i\Omega}\int dte^{-i\Omega t}<[\overrightarrow
{j}(t),\overrightarrow{j}(0)]> \label{sigdef}%
\end{equation}

At frequencies below interband transitions the current operator
$\overrightarrow{j}$ is  $\overrightarrow{j}=\partial\varepsilon
_{p}/\partial \overrightarrow{p}$. In the matrix notations of the preceding section,
the operator for current flow in the $x$ direction is

\begin{equation}
\mathbf{j}_{x} = \frac{\partial \varepsilon_p}{\partial p_x}
\label{jxop}%
\end{equation}
explicitly, 
\begin{equation}
j_{x,p}=\frac{\partial\varepsilon_{p}}{\partial p_{x}}=2t_{1}\sin p_{x}%
-4t_{2}\sin p_{x}\cos p_{y}+4t_{3}\sin2p_{x} \label{jx}%
\end{equation}
and we have chosen units such that the in-plane lattice constant is unity.

An important aspect of the conductivity is the spectral weight, or integrated
area, which is most conveniently expressed as an energy via
\begin{equation}
K(\Omega)=\frac{\hbar c}{e^{2}}\int_{0}^{\Omega}\frac{2d\omega}{\pi}\sigma
_{1}(\omega) \label{komegadef}%
\end{equation}
Here $c$ is the c-axis lattice parameter and $\hbar/e^{2}=4k\Omega$. Within
band theory the total kinetic energy associated with optical transitions
within the conduction band is
\begin{align}
K_{band}(\Omega &  =\infty)=\int\frac{d^{2}p}{\left(  2\pi\right)  ^{2}%
}f(\varepsilon_{p})\label{Kband}\\
&  \left(  2t_{1}\cos p_{x}-4t_{2}\cos p_{x}\cos p_{y}+8t_{3}\cos2p_{x}\right)
\nonumber\\
&  \approx0.43\text{ }eV
\end{align}
with less than $2\%$ doping dependence in the range $x=0.12-0.18$.
Interactions are expected \cite{Millis04d,Millis90,Schulz90,Stafford93} to reduce $K$ below its
band theory value; the amount of the reduction is a measure of the correlation strength
\cite{Millis04d}.

It is sometimes useful to express measured conductivities in terms of the
optical mass $m^{\ast}$ and scattering rate $\Gamma$ defined via%

\begin{equation}
-i\Omega\frac{m^{\ast}}{m}_{opt}+\Gamma_{opt}=\left(  \frac{\sigma_{1}(\Omega)+i\sigma
_{2}(\Omega)}{K_{band}}\right)  ^{-1} 
\label{mgamdef}%
\end{equation}

Note that we choose to normalize the conductivity to the band theory kinetic
energy, Eq \ref{Kband}.  A wide variety of other choices have been employed
in the literature; different choices lead to different over-all magnitudes for
the optically defined mass and scattering rate.  We normalize the data to 
$K_{band}$ because we are interested in the differences between measurements
and the predictions of band theory.

A general expression for the conductivity is complicated \cite{Millis03a}. If
one makes the assumption that 'vertex corrections' are negligible, then the
conductivity becomes $\sigma=\frac{e^{2}}{\hbar c}\frac{\Pi^{qp}}{i\Omega}$
(in 'imaginary time') with
\begin{align}
\Pi^{qp}(\Omega)  &  =T\sum_{n}\int\frac{d^{2}p}{\left(  2\pi\right)  ^{2}%
}\label{piqp}\\
&  \mathbf{j}_{x}(p)\operatorname{Im}G(p,i\omega
)\mathbf{j}_{x}(p)\operatorname{Im}G(p,i\omega+i\Omega)
\nonumber
\end{align}
and the Green function $G$ given as usual by%
\begin{equation}
G(p,i\omega)=\int_{0}^{1/T}d\tau e^{-i\omega\tau}<T_{\tau}d(\tau)d^{\dag}(0)>
\label{Gdef}%
\end{equation}

In particular, the dissipative conductivity is%
\begin{align}
\sigma_{1}^{qp}(\Omega)  &  =\frac{e^{2}}{\hbar c}\int_{-\infty}^{\infty}%
\frac{d\omega}{\pi}\frac{\left[  f(\omega)-f(\omega+\Omega)\right]  }{\Omega
}\label{sig1}\\
&  \int\frac{d^{2}p}{\left(  2\pi\right)  ^{2}}\mathbf{j}_{x}%
(p)\operatorname{Im}G(p,\omega)\mathbf{j}_{x}(p)\operatorname{Im}%
G(p,\omega+\Omega)\nonumber
\end{align}
$f$ is the Fermi function. Note that we have chosen units such that the
in-plane momentum is dimensionless as is the product $j_{x}G$.

\section{Electron Self Energy}

Doped Mott insulators, generally, and cuprate materials, in particular, 
appear \cite{Varma89} to be characterized by an electronic
self energy \ of unknown origin, which is not small and exhibits a significant
temperature and frequency dependence.  One widely used model is the "marginal
Fermi liquid" \cite{Varma89}. \ Another class of model self energies, with many
similar features, arises from 'spin Fermion models' for materials near
magnetic critical points \cite{Jiang96,Stojkovic97,Haslinger01}.  the 
single-site dynamical mean field method leads to a qualitatively similar
self energy. The
calculations reported below of the optical conductivity use a slightly
different self energy, of the form

\begin{equation}
\Sigma(\omega)=\gamma_{imp}+\Gamma(T)\left(  1-\lambda(T)\frac{\omega
_{c}(\omega_{c}+i\omega)}{\omega_{c}^{2}+\omega^{2}}\right)  +Z\omega
\label{sigma}%
\end{equation}

This form is chosen phenomenologically. It  represents a 
self energy with an imaginary part which  
is small at low frequency and large at high frequency,
and without noticeable momentum dependence. This latter
assumption is consistent with the photoemission data 
presented in the previous section.

\begin{figure}[ptb]
\includegraphics[width=3.0in]{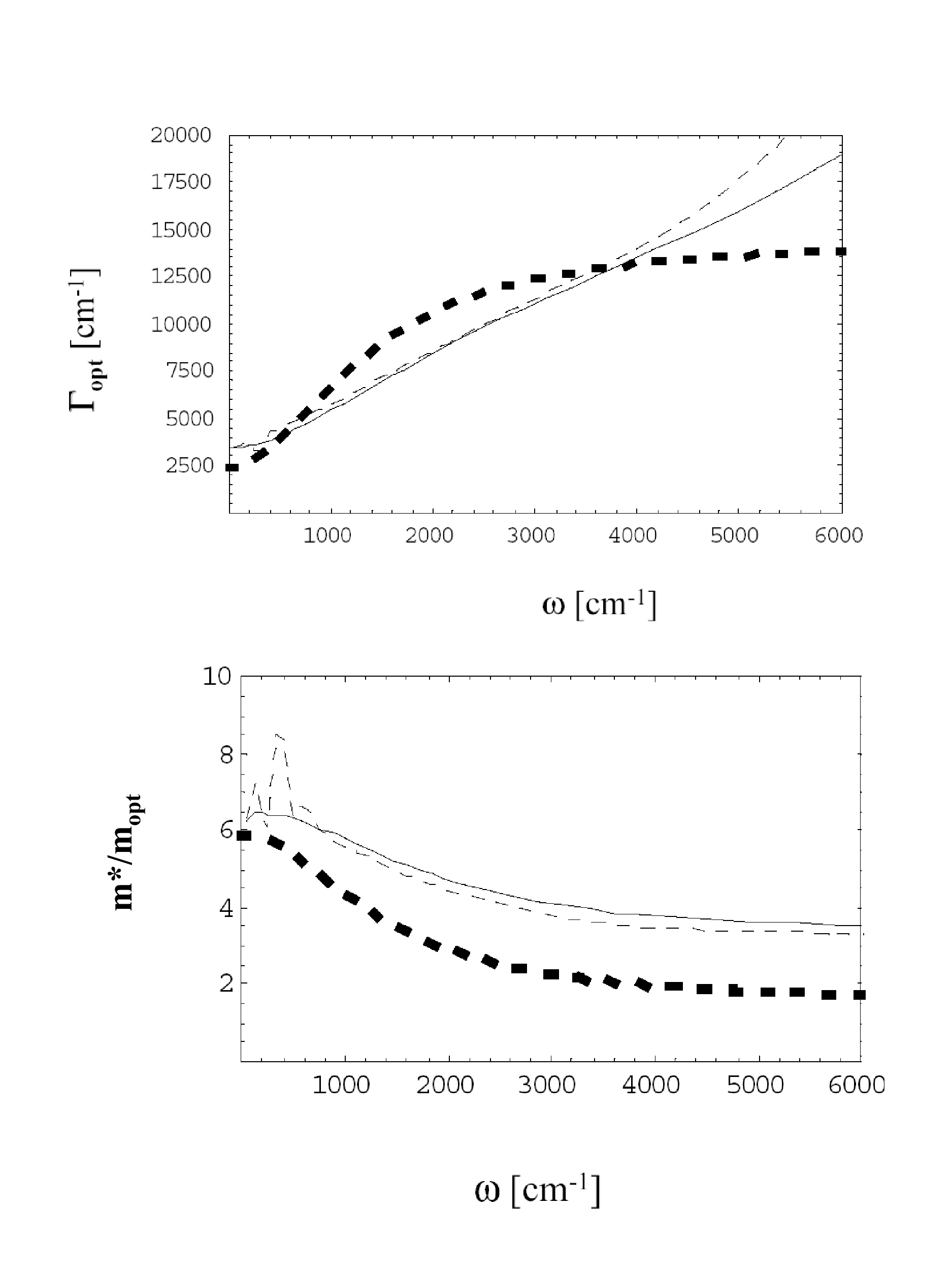}
\caption{\textit{{Scattering
rates (upper panel) and mass enhancements (lower panel) derived from room temperature 
optical  data for $x=0.17$ material, compared to results of
fits and to corresponding single particle rates.  Light dashed lines:
optically defined scattering rate $\Gamma_{opt}$ and mass enhancement
$m^*/m_{opt}$ obtained from optical
data using Eq \ref{mgamdef}. Light solid lines: 
$\Gamma_{opt}$ and $m^*/m_{opt}$
obtained by applying Eq \ref{mgamdef} to theoretical fit to optical conductivity.
Heavy dashed lines: 'single-particle' mass and
scattering rate obtained from model self energy used in fits to 
$x=0.17$ room temperature conductivity.}}}%
\label{fig:mandgamx17300}%
\end{figure}

Here $\gamma_{imp}$ is a constant scattering rate assumed to come from
impurities and the term $Z\omega$ expresses the effect of renormalizations
arising from physics at energies above the highest frequencies
considered in the analysis. The remaining 'many body' part of the self energy
is taken to be an inverted Lorentzian with frequency scale $\omega_{c}$, and
overall strength $\Gamma$. The parameter $\lambda(T)$ controls the crossover
from low to high frequency \ The key difference from the marginal Fermi liquid
form is the presence of a scale, $\omega_{c}$, which will be seen to be quite
low. The real and imaginary parts of the model self energy are shown as heavy
dashed lines in Figs \ref{fig:mandgamx17300} for the parameters used in
the calculation of the conductivity for $x=0.17$ (scattering rates extracted
from analysis of the optical conductivity are also shown; these will be
discussed below). The real part is displayed as a mass enhancement $m^{\ast
}/m=1-\operatorname{Re}\Sigma(\omega)/\omega.$

\section{Analysis of Data: Kinetic Energy and Mass Enhancement}

\subsection{Overview:}

\begin{figure}[ptb]
\includegraphics[width=3.0in]{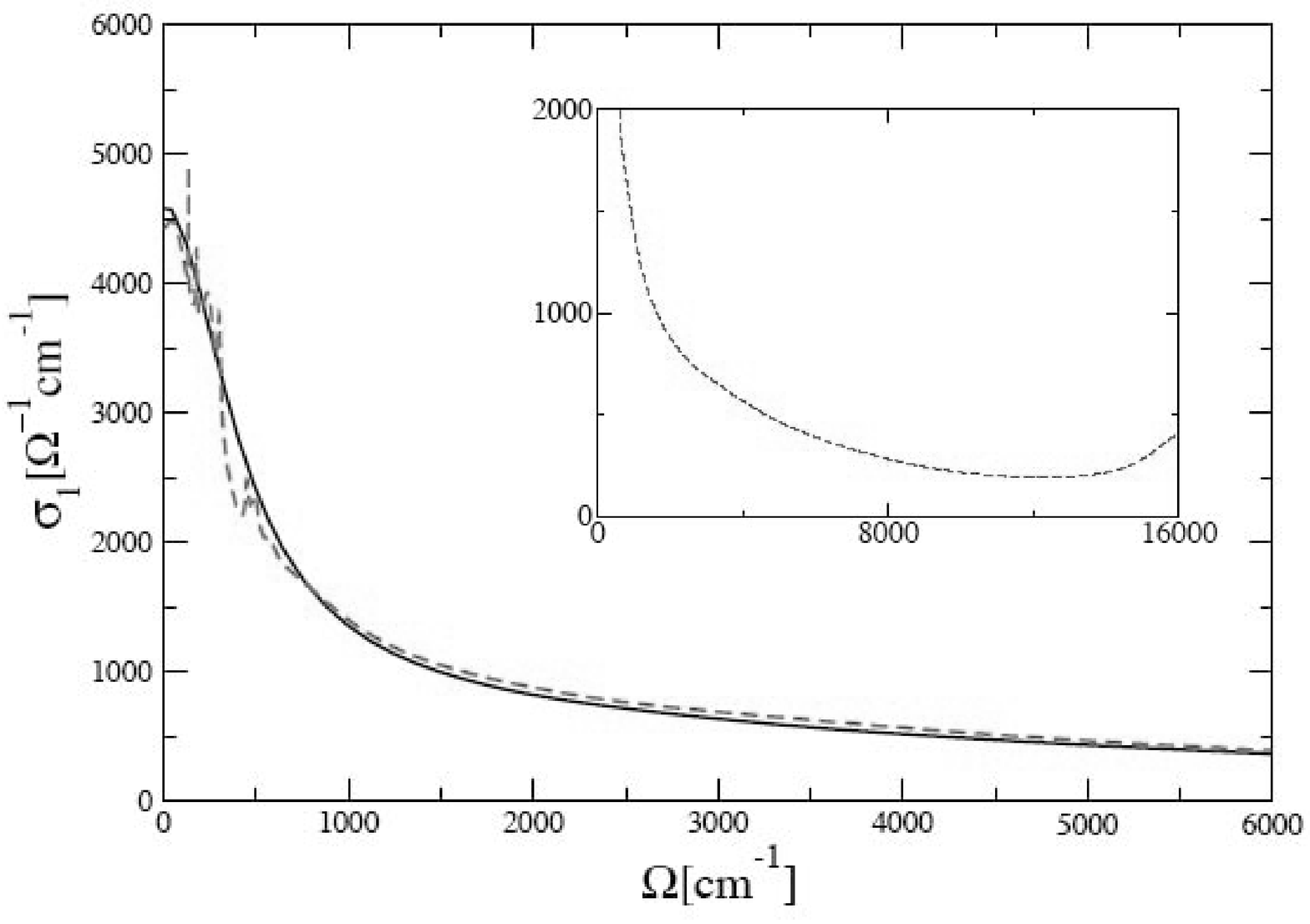}
\includegraphics[width=3.0in]{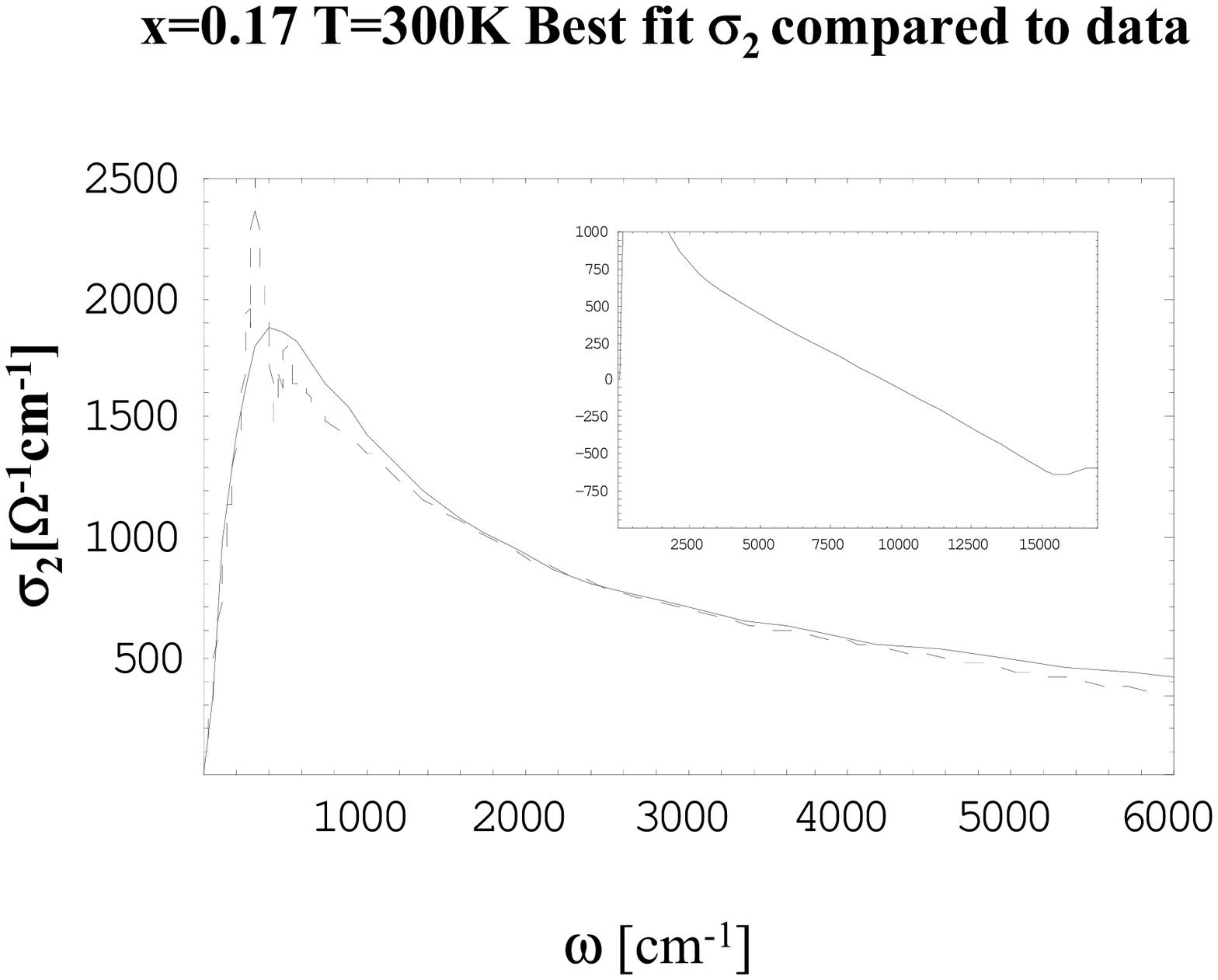}
\caption{\textit{{Measured
conductivities compared to theoretical model; $x=0.17$ and $T=300K$. Main
panels: measured dissipative (upper) conductivity and reactive (lower)
conductivities (dashed lines) compared to model calculation (solid lines).
Inset: measured conductivities over wide frequency range $0-2eV$.}}}%
\label{fig:x17sig300}%
\end{figure}

This section presents an analysis of important qualitative features of the
observed conductivity, in particular the kinetic energy (Eq \ref{komegadef})
and optical mass enhancement (Eq \ref{mgamdef}). The analysis relies
in an essential way on the assumption that in the frequency
range of interest both the real and imaginary
parts of the measured conductivity arise only from conduction band carriers,
with negligible effects of interband transitions. This assumption is
clearly correct at very low frequencies, and clearly breaks down at
sufficiently high frequencies. No clearly defined criterion
has appeared in the literature for estimating a frequency below
which the conductivity is dominated by the conduction band. 
The lack of a clear criterion for separating the
inter and intra-band contributions to the conductivity
is a source of systematic error whose magnitude is at present
unkonwn. The point
of view taken here is as follows.

The insulating end-member $Pr_2CuO_4$ has a conductivity characterized
by a gap at $\omega \approx 1.5eV$ \cite{Uchida91}. We believe that it is reasonable
to regard this gap as the Mott-Hubbard (or charge transfer)
gap and thus to understand 
the absorption at frequencies $\Omega \sim 1.5-2.5eV$ just above the 
gap  as being dominated by
the $CuO_2$ plane carriers of interest, while an increase at
higher frequencies indicates the onset of a strong interband transition. 
Modest (few percent)
doping destroys this large gap and redistributes the $1.5-2.5eV$ absorption  to
lower frequency \cite{Uchida91} while not changing the interband transitions
much. We therefore believe that at frequencies less than $1.5eV$ the observed
absorption comes from the $CuO_2$ plane carriers of interest. However,
an interband absorption at $\Omega >1.5eV$ will produce a contribution
to the reactive part of the conductivity at lower frequencies; thus only at frequencies
substantially less than $1.5eV$ will the total conductivity be dominated by
the conduction band carriers.

The insets of Fig \ref{fig:x17sig300} display the measured conductivities
of the $x=0.17$ sample
over a wide frequency range.  The reactive part $\sigma_{2}$ displays
a zero-crossing at about $\Omega \approx 1.25eV$; we believe
that this arises from the absorption feature visible at the high
end of the $\sigma_1$ frequency range.  We suggest that the 
real and imaginary conductivities are dominated by transitions
involving the $CuO_2$ plane carriers only
only at frequencies significantly less than $1eV$, and we will discuss the
data only for frequencies less than $\omega<0.75eV$, and the our main results
rely only on frequencies rather less than $0.5eV$.  As shall be seen below,
in this frequency range the analysis yields a consistent picture with conclusions
which are insensitive to the precise value of the upper cutoff frequency.

\subsection{Kinetic Energy}

The kinetic energy defined in Eq \ref{komegadef} is a fundamental measure of
the strength of interactions. If the cuprates were well described by band
theory, the low frequency optical conductivity would consist of a narrow
'Drude' peak concentrated at $\Omega=0$, with area leading to a kinetic energy
$K(\Omega)$ which would rapidly approach the value $K_{band}=0.43eV$ as
$\Omega$ increases from $0$.

\begin{figure}[ptb]
\includegraphics[width=3.0in]{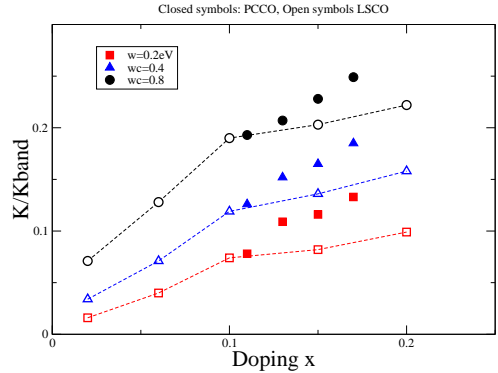} 
\caption{Solid symbols: kinetic
energy, obtained by integrating observed room temperature conductivity
for $Pr_{2-x}Ce_xCuO_4$ films
 up to several cutoff frequencies and normalizing to  the band theory value,  plotted against
doping. Open symbols: same quantity, obtained from published \cite{Uchida91}
data on hole-doped $La_{2-x}Sr_{x}CuO_{4}$ }%
\label{fig:kebothofx}%
\end{figure}

The filled symbols in fig \ref{fig:kebothofx} show the kinetic energy obtained
from the integral of the room temperature conductivity from zero to a cutoff
frequency $\Omega_c$,  normalized to the band value and plotted
against doping for different cutoff frequencies. One sees that for all dopings
and all relevant frequencies the kinetic energy is a very small fraction of the band
value. A linear doping dependence is evident, with slope approximately
independent of the cutoff frequency. The linear $x$ dependence (with slope of the
order of unity) and the small intercept are hallmarks of strong correlation 
or Mott physics \cite{Schulz90,Millis90,Stafford93,Millis04d}. 

The figure also shows as open symbols the 
same quantity obtained from published data \cite{Uchida91} for the hole-doped
material $La_{2-x}Sr_{x}CuO_{4}$, where a wider range of dopings are
available. One sees that the qualitative features of
strong reduction in magnitude and strong doping dependence occur in both 
electron-doped and hole-doped
materials. 
Interestingly,  while the  $La_{2-x}Sr_{x}CuO_{4}$ data at low dopings ($x\le 0.1$)
appears as a continuation of the doping dependence 
observed in $Pr_{2-x}Ce_xCuO_4$, in the hole doped material the
kinetic energy curves exhibit a pronounced break at about $x=0.1$. The break
in slope is not understood and does not, to our knowledge, follow
from any theory.  Extension to lower  
dopings  of measurements on the electron-doped compounds would be
very desirable, as would an improved understanding of the break in x-dependence
observed in the hole doped materials.    

The reasonable correspondences of the kinetic energy magnitudes
and doping dependences  suggests that
the electron doped materials are approximately as strongly
correlated as the hole doped ones. The approximately linear
doping dependence of $K$ suggests that the $U$ value
is not strongly doping dependent within the electron-doped
family of materials.  Both of these observations
are in apparent disagreement with recent theoretical studies of 
electron doped compounds \cite{Kusko02,Tremblay04} suggesting
that despite the evident successes of these theories in accounting
for the photoemission data, some issues remain in need of clarification.

\begin{figure}[ptb]
\includegraphics[width=3.0in]{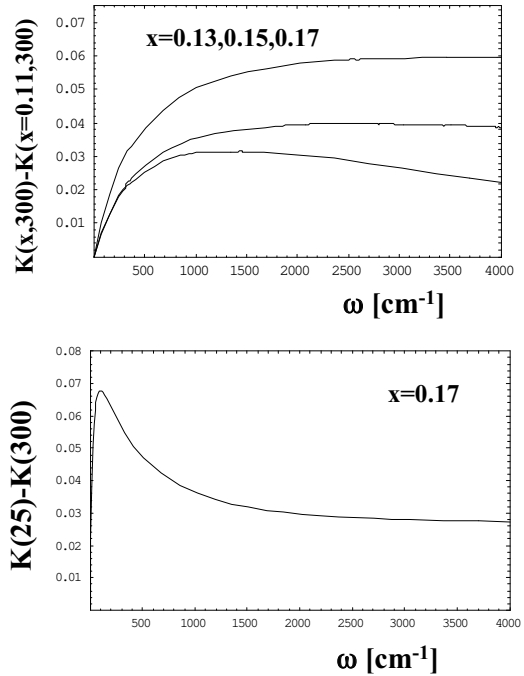} 
\caption{\textit{{Upper
panel: }} Frequency dependence of difference in room temperature kinetic
energy between $x=0.11$ sample and $x=0.17$ (upper curve) $x=0.15$ (middle
curve) and $x=0.13$ lowest curve. \textit{{Lower panel: }} Frequency
dependence of difference of room temperature and $T=25K$ kinetic energy of
$x=0.17$ sample. }%
\label{fig:deltakeofxandt}%
\end{figure}

As doping is increased the optical conductivity increases, but the increase is
not uniform in frequency. The upper panel of Fig \ref{fig:deltakeofxandt}
shows the room temperature kinetic energy difference $K(x,\Omega
)-K(x=0.11,\Omega)$ for $x=0.13,$ $0.15,$ $0.17$.  A rapid rise at low
frequency is evident, as is an approximate saturation at frequencies greater
than about $0.15eV$; in other words, the doped carriers contribute most
strongly to the low frequency conductivity. The kinetic energy is also
temperature dependent, increasing as $T$ is decreased. The lower panel of Fig
\ref{fig:deltakeofxandt} shows the changes in the measured kinetic energy of
the $x=0.17$ sample as temperature is varied between room temperature and
$25K$.  As the temperature is varied, two effects may occur: a redistribution
of spectral weight, arising because scattering rates and mass enhancements
have temperature dependences, and a change in the total conduction band spectral 
weight. Both effects are visible in the lower panel of Fig. \ref{fig:deltakeofxandt}: 
the sharp peak near zero frequency arises in part from a decrease in
the scattering rate, which narrows the 'Drude peak' leading to a pile-up of
spectral weight at low frequency. The saturation at higher frequencies shows
that in addition to the rearrangement, there is a net temperature dependent
increase, and that this increase affects mainly the low frequency conductivity
(since the difference curve is flat above $\omega \approx .1eV$.
The change in spectral weight as $T$ is decreased from room
temperature to $25K$ in the $x=0.17$ sample is seen to be of about the 
same magnitude as
the increase in spectral weight as doping is increased from $x=0.11$ to $0.17$

\subsection{Mass and scattering rate}

The dashed curves in Fig. \ref{fig:mandgamx17300} show the mass and
scattering rate computed by applying Eq \ref{mgamdef} to the $x=0.17$ room
temperature data for frequencies up to about $6000$ $cm^{-1}$ $=0.75eV$. 
The light solid curves are obtained from a theoretical fit 
to $\sigma$ discussed in detail in the next section.
The upturn in the experimental optical scattering rate beginning at about 
$6000$ $cm^{-1} \approx 0.75eV$ is caused mathematically 
by the decrease of $\sigma_2$  towards its $1.25eV$ zero crossing.
In physical terms, this is a signature that the data at $\omega \gtrsim 0.75eV$
are significantly affected by an interband transition,
rendering an interpretation in terms of
mass and scattering rate meaningless. The smooth behavior observed 
at lower frequencies suggests that  for frequencies below
about $1/2eV$ an interpretation in terms of a single band characterized by
a mass and scattering rate is reasonable--however, it is possible that even in this
regime, interband effects are important. We will return to this issue in the conclusion. 
The $\omega <0.5eV$ 
data are consistent with an optical scattering rate which is reasonably linear
in frequency, as in the hole-doped materials \cite{Schlesinger89,Orenstein90}.

One sees from Fig. \ref{fig:mandgamx17300} that the optical mass
enhancement (relative to band theory) is large, of order 8, even at room
temperature, and is strongly frequency dependent. The large mass is simply the
restatement of the suppression of kinetic energy discussed above. The
characteristic scale of the frequency dependence seen directly in the
crossover of the optical mass from its low frequency to its high frequency
value is of the order of $0.25eV$.  The high and low frequency masses
are seen to differ by about a factor of two. One may therefore infer that
two processes are at work: an over-all suppression of spectral weight
(characterized by a frequency scale higher than the highest frequency
we analyse) and an additional lower frequency ($\omega <0.25eV$)
effect which enhances the mass by an additional factor of about 2.
Analysis (not shown) of other dopings reveals almost identical behavior.

\section{Modelling of data}

\subsection{Overview}

This section considers the modelling of the optical data within the
no-vertex-corrections approximation. It begins with an analysis of the
$x=0.17$ compound, in which no signatures of density wave order are visible.
The next subsection concerns the room temperature behavior as a function of
doping (where again no signatures of a density wave gap are evident except
perhaps in the $x=0.11$ sample), and the final subsection deals with the
effect of a density wave gap on the conductivity.

\subsection{x=0.17}

This section discusses the modelling of the $x=0.17$ optical data. A
fundamental assumption is that the conductivity at scales less than $1eV$ is
described by a single band of carriers. The insets in the two panels of Fig
\ref{fig:x17sig300} suggest that interband transitions become important only
above about $2eV$. \ The kinetic energy analysis of the previous section shows
that the oscillator strength in the $\omega<1eV$ conductivity is substantially
less than is predicted by band theory. An analysis involving strong
correlations is therefore needed. Here it will be assumed, consistent with the
'single-site dynamical mean field approximation' \cite{Georges96} and with
many other works \cite{Varma89,Jiang96,Stojkovic97,Haslinger01}, that the main
effect of the correlations is to produce an electron self energy which may be
large and strongly frequency dependent. \ 

The solid lines in Fig. \ref{fig:x17sig300} show the results of theoretical
calculations based on Eqs \ref{sig1} and \ref{sigdef}, with
$\Gamma=0.9eV,$ $\lambda=0.83,$ $Z=0.5$ and $\omega_{c}=0.17eV$. \ The
agreement with data (dashed lines) is reasonable. The real and imaginary parts
of the electronic self energy are shown as heavy dashed lines in Fig.
\ref{fig:mandgamx17300}. The self energy is characterized by a
surprisingly low frequency scale (0.17eV) and by a very large magnitude. The
high frequency limit of the
imaginary part of the self energy is found to be frequency independent,
with a value $\approx 1.5$
$eV$, comparable to the band width. The low frequency mass enhancement is
correspondingly large (of order $6$), and arises mostly from processes acting
at the relatively low scale set by $\omega_{c}$. \ The fits involve a function
of several parameters, and it is therefore difficult to say precisely what are
the uncertainties in the resulting values. However, $\Gamma$ sets the rate at
which the $\omega>3000cm^{-1}$ conductivity drops with frequency, $Z$ sets the
magnitude of the conductivity in the high frequency regime, $\lambda$ sets the
value at low $\omega$, and $\omega_{c}$ determines the scale at which the
behavior crosses over from low to high frequencies. Significant (more than
$10\%$) changes in any one of these parameters leads to noticeable decreases
in the quality of the fits.

It is evident from Fig \ref{fig:mandgamx17300} that the optically defined
mass and scattering rates are not faithful representations of the single
particle mass and scattering rates. In particular the calculated optical
scattering rate is much more nearly linear than is the imaginary part of the
model single-particle self energy, while the assumed single particle mass
drops off more quickly than the optically defined one.

\subsection{Doping dependence}

\begin{figure}[ptb]
\includegraphics[width=3.0in]{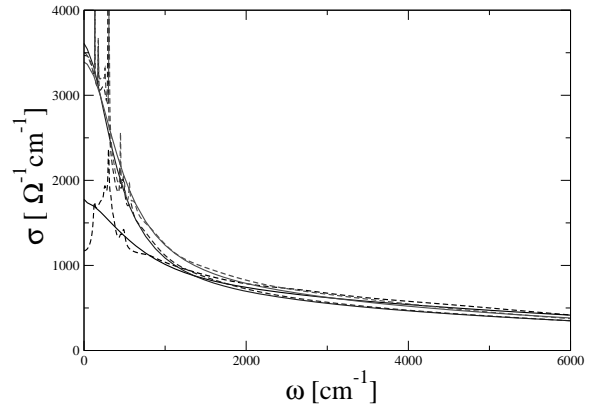}
\caption{\textit{Real part of
measured room temperature conductivity (dashed lines) compared to theoretical
model (solid lines) for $x=0.11$, $0.13$, $0.15$.}}%
\label{fig:sigx111315}%
\end{figure}

\begin{figure}[ptb]
\includegraphics[width=3.0in]{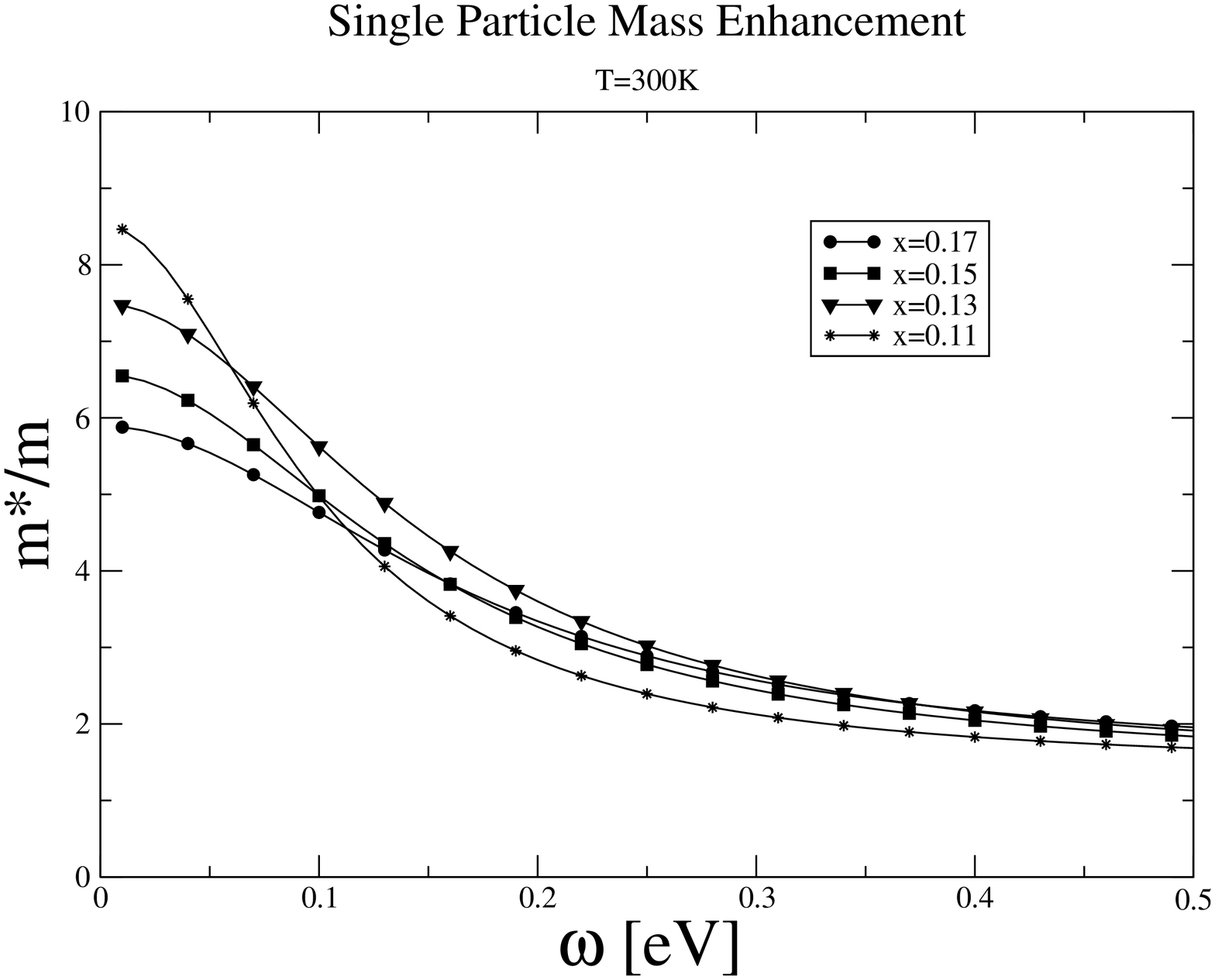}
\caption{\textit{Frequency dependent
single particle mass enhancement $m^{*}/m=1-Re\Sigma(\omega)/\omega$ inferred
from self energy only analysis of room temperature conductivity.}}%
\label{fig:msingle}%
\end{figure}

This subsection summarizes the results of fitting the room temperature
conductivities with the quasiparticle-only model. Fig \ref{fig:sigx111315} shows
the measured room temperature conductivities and the best-fit calculated
conductivities (solid lines) at the other available dopings $x=0.11,0.13,0.15$. 
Fig \ref{fig:msingle} shows the single-particle mass
enhancement inferred from these fits. The parameters are indicated in 
Table \ref{Table}.  

Fig. \ref{fig:msingle} shows that
for $x=0.13$, $0.15$ and $0.17$ there is a negligible doping dependence of
$m^{\ast}$ at high frequencies while there is a characteristic, low
frequency scale, below which the mass sharply increases, in a manner which
depends on doping. This is a restatement, in the language of single particle mass, of the
observation made above that the effect of doping is to add spectral weight at
low frequencies.  

The $x=0.11$ sample is seen to deviate from the monotonic behavior. We
will argue below and in a companion paper 
that this deviation is associated with the presence of
antiferromagnetism even at room temperature.

For all dopings, the zero frequency
(Fermi surface) mass enhancements implied by the analysis are very
large. As seen in the Table, this implies
quasiparticle velocities substantially suppressed relative to band velocities.

\textit{}

\begin{table}[htdp]
\caption{Self energy parameters and associated velocity renormalization implied
by self-energy-only analysis of room temperature conductivity}
\begin{center}
\begin{tabular}
[c]{cccccc}%
Doping & $\Gamma\lbrack eV]$ & $\lambda$ & $\omega_{c}[eV]$ & $Z$ & $v^{\ast
}/v_{band}$\\
0.11 & 0.95 & 0.75 & 0.1 & 0.41 & 0.12\\
0.13 & 1.1 & .83 & 0.15 & 0.41 & 0.13\\
0.15 & 0.98 & 0.79 & 0.15 & 0.41 & 0.15\\
0.17 & 0.9 & 0.83 & 0.17 & 0.5 & 0.17
\end{tabular}
\end{center}
\label{Table}
\end{table}

\subsection{Low Temperature analysis}

\begin{figure}[ptb]
\includegraphics[width=3.0in]{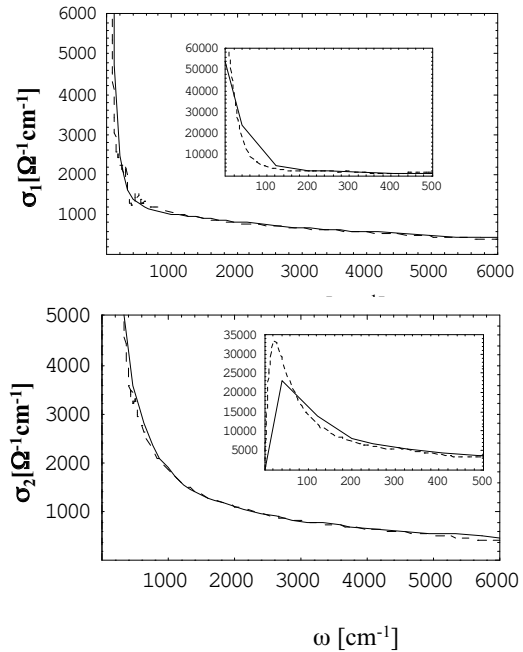}
\caption{\textit{{Main
panels: measured dissipative (upper) conductivity and reactive (lower)
conductivities (dashed lines) compared to model calculation (solid lines)
for $x=0.17$ and $T=25K$.
Inset: expanded view of low frequency regime}}}%
\label{fig:x17sig25}%
\end{figure}

\begin{figure}[ptb]
\includegraphics[width=3.0in]{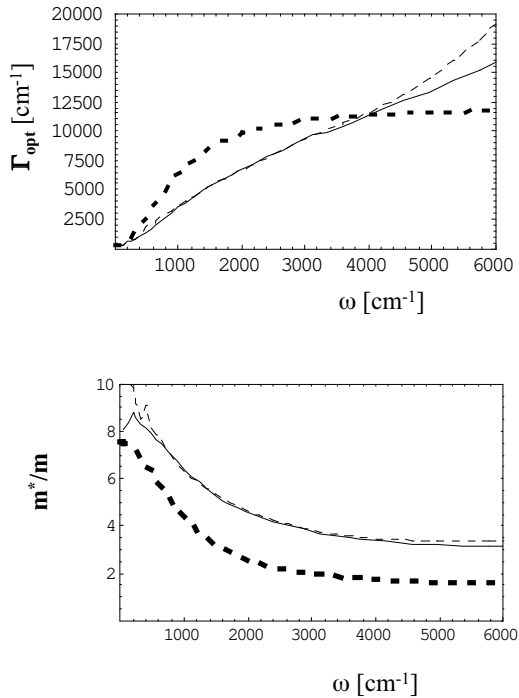}
\caption{\textit{{$T=25K$
scattering rate (upper panel) and mass enhancement (lower panel)} derived from
optical data (light dashed line), from theoretical fits to optical data for the $x=0.17$
sample (light
solid line) compared to 'single-particle' mass and scattering rate obtained
from model self energy (heavy dashed lines).}}
\label{fig:mandgamx1725}
\end{figure}

Figs \ref{fig:x17sig25} show data and model conductivities at low temperature
($25K$). Again the agreement is reasonably good, although as can be seen from
the inset the model underestimates the mass enhancement and overestimates the
scattering rate at low frequencies. \ The 'best fit' parameters for the low
frequency data are $\Gamma=0.75eV,$ $\lambda=0.98,$ $Z=0.4$
and $\omega_{c}=0.12eV$. \ \ The decrease in $Z$ reflects the increase in
conduction band kinetic energy; the increase in $\lambda$ reflects the smaller
value of the $dc$ scattering rate; the decrease in $\Gamma$ reflects the
smaller value of the high frequency optical scattering rate (more rapid
decrease of $\sigma_{1})$ and the decrease in $\omega_{c}$ reflects the lower
frequency and more rapid crossover of the data from high to low frequency.
\ The actual value of the high frequency conductivity has very little
temperature dependence; thus in this parametrization changes in the parameters
$Z$ and $\Gamma$ controlling the high frequency conductivity must compensate
each other. Figs. \ref{fig:mandgamx1725} show the single particle and
optically defined mass and scattering rate following from the fits to the low
temperature data. The qualitative behavior is very similar to that of the
higher temperature results.

\begin{figure}[ptb]
\includegraphics[width=3.0in]{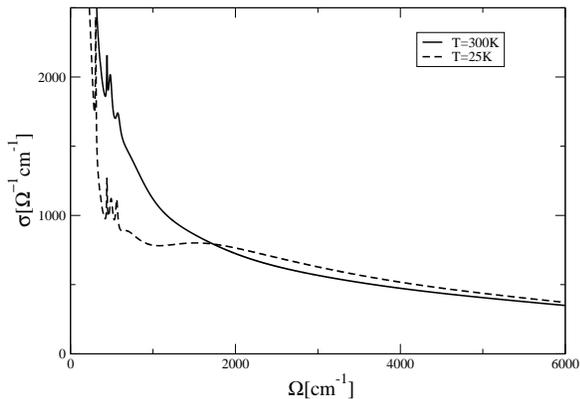}
\caption{\textit{Measured conductivity for
$x=0.13$ sample at room temperature and low temperature.}}
\label{fig:x13}
\end{figure}

Difficulties arise in applying this analysis to the low $T$ behavior
of the  lower $x$ samples. The
issue is most clearly revealed by examination of the data for $x=0.13$ sample.
Comparisons of the dashed and dotted lines in Fig. \ref{fig:x13} shows that as
temperature is decreased the dissipative conductivity does not simply shift
towards lower frequency, as does the conductivity of the $x=0.17$ material.
Instead, an upward shift occurs in frequency, as expected if a density wave gap
opens up. The conductivity of this material will be discussed 
in terms of antiferromagnetism in a companion paper.

\section{Comparison to Photoemission}

\begin{figure}[ptb]
\includegraphics[width=3.0in]{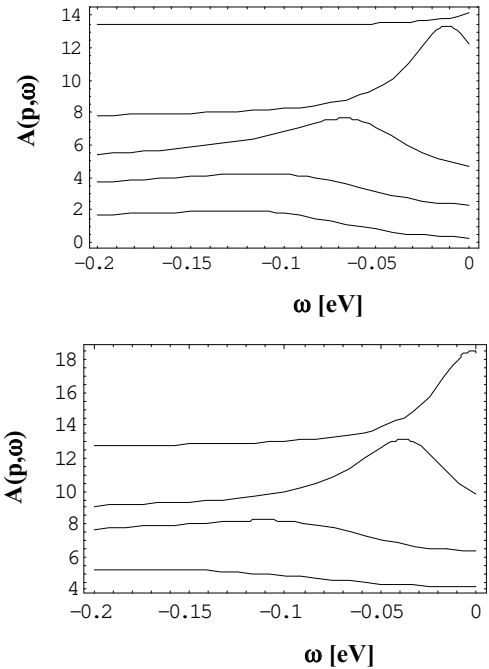} 
\caption{\textit{{Energy dispersion
curves (electron spectral function as function of energy for fixed momentum)
calculated for $300K$ best-fit optical parameters. Top panel: momenta along
zone face $p=(\pi,(0,0.1,0.2,0.3,0.4)*\pi)$ Lower panel: zone diagonal,
$p=0.4,0.6,0.8,0.9*\pi,\pi)$.}}}%
\label{fig:x017edc}%
\end{figure}

The parameters obtained from the fits to the optical data make predictions
for photoemission spectra. To facilitate comparison to experiment
we  plot these as 
energy dispersion curves
('EDC') which are the imaginary part of the electron Green function, plotted as
a function of energy at fixed momentum. The two panels of Fig.
\ref{fig:x017edc} show EDC traces calculated using the self energy which best
fits the room temperature $x=0.17$ optics calculation.

The 'no-vertex-corrections' modelling of the conductivity requires that the
quasiparticle mass enhancement be very large in order to account for the
suppression of spectral weight relative to band theory. The predicted Fermi
velocity renormalizations are given in the last column of the Table. These
imply zone-diagonal Fermi velocities of order $0.5eV-\mathring{A}$, as may be
seen directly from the dispersions of the peaks in Fig. \ref{fig:x017edc}. The
velocities have been directly measured in photoemission experiments
\cite{Armitage02} and are found to be much larger, of order \thinspace
$1.5-2eV-\mathring{A}$, as may be seen in Fig \ref{fig:dispersion}.
This dramatic inconsistency shows that the
self-energy-only model is not applicable to the electron-doped cuprates. In
Fermi-liquid terms, a rather large vertex correction is important. Theories of
the conductivity \cite{Abrahams01,Stojkovic97,Haslinger01} which do not
include this factor seem unlikely to be directly applicable to the actual materials.

\section{Conclusion}

We have presented a detailed analysis of the optical conductivity of a series
of films of electron-doped cuprates. We have shown, via an analysis of the
kinetic energy, that the materials are approximately as strongly correlated as
the corresponding hole-doped systems. We have performed detailed
modelling of the observed optical spectrum  using a self energy which depends
on frequency only.  The essential assumption underlying this analysis
is that at relevant frequencies  the real and imaginary
parts of the measured conductivity are dominated by the response of 
mobile carriers moving in the band structure inferred from band theory
calculations and scattered by some frequency and temperature
dependent scattering mechanism. The optical response at low
frequencies corresponds to a large (factor $6-10$) mass enhancement 
relative to band theory; within the scattering only model this mass enhancement
would imply a photoemission spectrum characterized by a velocity much
lower than is actually observed.
Thus  we conclude that the doping dependent
suppression of optical oscillator strength cannot be due solely to local physics of
the sort considered in the single-site dynamical mean field theories of doped
Mott insulators or in a variety of phenomenological models.  
As an aside we find that the scattering rate inferred from an 'extended Drude'
analysis of the conductivity need not be a particularly faithful representation of the
underlying single particle scattering rate (see Fig. \ref{fig:mandgamx17300}).

The modelling was performed on the assumption that at frequencies
less than $\approx 0.5eV$ the real and imaginary parts of the measured conductivity
arise from scattered conduction band carriers. For frequencies greater than
about $0.75eV$ this assumption was shown to be untenable: a high frequency
interband transition produces a negative contribution to $\sigma_2$, leading
to an unphysical upturn in the scattering rate (see Fig.  \ref{fig:mandgamx17300}
and section IV-C).  Although we presented qualitative
arguments that at frequencies less than $1eV$ the absorbtive part of the conductivity 
is due to conduction band carriers, these arguments are not conclusive. it is
possible that even at lower frequencies, interband transitions could appear,
invalidating the analysis presented here. If this occurred, some of the optical
oscillator strength presently assigned to the conduction band would be reassigned
to an irrelevant transition; the optical masses would therefore be increased,
worsening the discrepancy between optical and photoemission masses.
We therefore suspect that our main finding, of a large discrepancy between
the predictions of a self-energy-only theory and the photoemission and optical
data, is robust.

Taken together, the available data present the following conundrum. Optical spectral
weight is proportional to a carrier density times a charge squared 
divided by a carrier mass: $K \sim  \frac{n_{eff}e_{eff}^2}{m_{eff}}$.
Two conventional interpretations of a small spectral weight are a small
number of carriers or a large carrier mass. The photoemission
measurements reveal (at least for large dopings where density wave effects are
absent) a large Fermi surface, consistent with band theory, and ruling out
a simple small carrier number picture.  Hall effect measurements \cite{Dagan04}
also indicate for optimally doped materials 
a carrier number reasonably consistent with the band theory value. A large body
of theoretical work \cite{Varma89,Georges96,Jiang96,Stojkovic97,Haslinger01} 
relates the optical conductivity to an essentially local scattering rate and the 
spectral weight suppression to an enhanced quasiparticle mass. Our analysis
shows that these theories require a mass enhancement larger by a factor
of at least 4 than is directly observed in photoemission.  One is therefore forced
to look to  a renormalization of
the 'effective charge'. In Fermi liquid language, this renormalization is expressed as
a vertex correction or Landau parameter, so one   must assume
that the Mott correlations are expressed by a vertex correction
which diverges as the doping decreases. Unfortunately, 
the most straightforward implementation of the vertex
correction idea  strongly disagrees with measurements of
the magnitude of the superconducting penetration depth (\cite{Ioffe02a}). 
Construction of a viable theory of the optical conductivity of this
(and perhaps other) strongly correlated two dimensional materials
remains an important open problem.

\textit{Acknowledgements: } AJM acknowledges  support from NSF-DMR-0431350 and the
CNRS.

\end{document}